\documentclass[fleqn,a4paper,12pt]{article}
\usepackage{amssymb}
\usepackage{amsmath}
\usepackage[dvips]{graphicx}
\usepackage{placeins}
\usepackage{lscape}
\usepackage{a4}
\usepackage{epsfig}

\newcommand{\ri}{{\mathrm i}}
\newcommand{\p}{\partial}
\newcommand{\bea}{\begin{array}}
\newcommand{\eea}{\end{array}}

\catcode`@=11 \long
\def\@caption#1[#2]#3{\par\addcontentsline{\csname
ext@#1\endcsname}{#1} {\protect\numberline{\csname
the#1\endcsname}{\ignorespaces #2}} \begingroup \small
\@parboxrestore \@makecaption{\csname fnum@#1\endcsname}
{\ignorespaces #3}\par \endgroup} \catcode`@=12

\newcommand{\Q}{\mathbb{Q}}

\newcommand{\la}{\label}

\catcode`@=11 \long
\def\@caption#1[#2]#3{\par\addcontentsline{\csname
ext@#1\endcsname}{#1} {\protect\numberline{\csname
the#1\endcsname}{\ignorespaces #2}} \begingroup \small
\@parboxrestore \@makecaption{\csname fnum@#1\endcsname}
{\ignorespaces #3}\par \endgroup} \catcode`@=12

\begin{document}

\allowdisplaybreaks
 \begin{titlepage} \vskip 2cm

\begin{center} {\Large\bf Superintegrable systems with spin invariant w.r.t. the rotation group}
\footnote{E-mail:
{\tt nikitin@imath.kiev.ua} } \vskip 3cm {\bf {A. G. Nikitin }
\vskip 5pt {\sl Institute of Mathematics, National Academy of
Sciences of Ukraine,\\ 3 Tereshchenkivs'ka Street, Kyiv-4, Ukraine,
01601\\}}\end{center}
\vskip .5cm \rm
\begin{abstract}Quantum nonrelativistic systems with $2\times2$
matrix potentials are investigated. Physically, they simulate
charged or neutral fermions with  non-trivial dipole momenta,
interacting with an external electric field. Assuming rotationally
invariance of the Hamiltonian all such systems allowing second order
integrals of motion  are identified.  It is shown that the integrals
of motion can be effectively used to separate variables and to reduce the systems to decoupled ordinary differential equations.
Solutions for two of the discussed problems  are presented explicitly.
\end{abstract}
\end{titlepage}
\section{Introduction\label{intro}} Exactly solvable systems of quantum mechanics are favourite subjects for many physicists and mathematicians. The beauty of such systems (like the Hydrogen atom or the harmonic oscillator) is that they are simple enough to be solved in a way free of uncertainties and inconveniences of various approximate approaches. Meanwhile, they are sufficiently complicated to model the physical reality. In addition, the complete sets of their exact solutions supply us by convenient bases for expansions of solutions of other problems. Many of exactly solvable  systems admit nice hidden symmetries  which are very interesting on their own account.

  The exact solvability of quantum mechanical systems is usually caused by their specific property called superintegrability. The system with $n$ degrees of freedom  is integrable if it admits $n-1$ commuting integrals of motion in addition to its  Hamiltonian. The system is superintegrable if it is integrable and admits  at least one more integral of motion.

The systematic search for superintegrable systems started with paper \cite{wint1}. We will not recount the details of rather inspiring history of this search and discuss all obtained fundamental results related to quantum mechanical and classical scalar systems. In contrary, we restrict ourselves to discussion of superintegrable systems with spin.

Such systems where studied methodically in recent papers \cite{w6}-\cite{w8}. The problem of classification of superintegrable systems with spin was  stated in \cite{w6} where 2d systems with spin-orbit interaction and first order integrals of motion had been presented. Then these results were generalized to the cases of three-dimensional Euclidean space \cite{w7} and second order integrals of motion \cite{w8}. The results presented in \cite{w8} were restricted to rotationally invariant systems and to integrals of motion which are rotational vectors or scalars.

However, the spin-orbit coupling is not the only spin effect which can be present in quantum mechanical systems. One more coupling which is very important and has a  well observable effect is the dipole, or Pauli interaction. This interaction is represented by the Stern-Gerlach term  $\sim {\bf S}\cdot{\bf B}$ (where $\bf S$ and $\bf B$ are the spin and magnetic field strength vectors) ore, more generally, by a matrix term linear in an external field. Moreover, the Pauli interaction affects  even  neutral particles provided, like neutron, they have  non-trivial dipole moments.

An important example of a 2d superintegrable system with dipole spin
interaction was presented long time ago by Pron'ko and Stroganov \cite{Pron} . However, a systematical search for such systems is only at the beginning. The classification of 2d systems with the first order integrals of motion was presented in \cite{N1}, while the 3d systems with Fock type dynamical symmetry were derived in \cite{N2}. Generalizations of the Pron'ko-Stroganov system to the case of arbitrary spin were discussed in \cite{Pron2} and \cite{N6}. However, a classification of integrable and superintegrable systems with the dipole interaction, admitting higher order integrals of motion, was not carried out till now.

In the present paper 3d superintegrable systems with dipole interaction are classified. We restrict ourselves to systems which are invariant w.r.t. the rotation group and admit second order integrals of motion. The list of such systems is not too long, but two  of them are defined up to arbitrary functions, and so the number of inequivalent systems is infinite. In particular, it includes a system with  Fock symmetry discussed recently in \cite{N2}, and a supersymmetric system with a matrix potential. For a classification of supersymmetric matrix potentials see papers \cite{N3}--\cite{K}.

Superintegrable systems are in many cases  exactly solvable, see, e.g., \cite{Ka}-\cite{F}. In particular, all the spinless 2d superintegrable systems  whose integrals are
given by second order differential operators are exactly solvable as well, and it was conjectured  that this property is valid also for higher
dimensional superintegrable systems \cite{w1}.

For systems with spin the situation is  more complicated due to the presence of the additional dichotomous variable. As it was noticed in \cite{N2}, such systems with 3 spatial variables are not necessary exactly solvable even if they admit more than 3  second order integrals of motion.

Thus the solvability of the discussed  systems should be examined separately.
Of course, it is impossible to solve all systems with arbitrary functions, presented in  the current paper. We restrict ourselves to two particular cases which are seemed to be physically interesting since  include the field of point charge.
The corresponding exact solutions are presented in section 6, where some other exactly solvable problems are indicated also. In addition, we apply integrals of motion to reduce generic (i.e., including arbitrary functions) eigenvalue problems to decoupled systems
of ordinary differential equations for radial wave functions. However, these equations in general are not exactly solvable.
\section{Schr\"odinger-Pauli equations for neutral particles}

To describe spin effects in non-relativistic quantum mechanics
the Schr\"odinger equation should be generalized by Pauli
term proportional to the scalar product of spin with vector
of the magnetic field strength.
For neutral particles with non-trivial dipole moments
(e.g., for neutrons) this term becomes dominant  since in this case
the minimal interaction is absent. The Pauli-like term is requested
also for description of interaction of charged  particles having
non-trivial electric moments with an external electric field.

In other words, there are several reasons to study the Schr\"odinger-Pauli
equations of the following generic form:
\begin{gather}\la{SP} H\psi\equiv\left(\frac{p^2}{2m}+\frac\lambda{2m}\mbox{\boldmath
$\sigma$}\cdot{\bf K}+\omega V({\bf x})\right)\psi=E\psi\end{gather}
where  $\mbox{\boldmath $\sigma$} $ is the matrix vector whose
components are Pauli matrices, ${\bf K}$ and $V$ are vector and
scalar external fields. Moreover, $\lambda$ and $\omega$ are
coupling constants and $E$ denotes an eigenvalue of Hamiltonian $H$.

Since we are interested in systems with a spin dependent interaction, constant $\lambda$ is supposed to be nonzero. To obtain more compact formulae in the following calculations let us rescall  variables and reduce the
Hamiltonian  to the following simplified form:
\begin{gather} H=-\Delta+\hat V({\bf x})\equiv-\Delta+\mbox{\boldmath $\sigma$}\cdot {\bf F}
+F^0\label{Redu}\end{gather} where $\Delta$ is the Laplace operator and $\hat V({\bf x})$ is a {\it matrix potential} .
For this purpose we change
 in (\ref{SP}) \begin{gather}\label{change}E\to{\hat E}={2m}E, \quad {\bf K} \to {\bf F}=
\lambda {\bf K},\quad V
\to F^0={2m\omega}V.\end{gather}

Just Hamiltonians (\ref{Redu}) which include the generic $2\times2$
matrix potential $\hat V$ will be the subject of our classification.
Our goal is to find all possible external fields ${\bf
F}=(F^1,F^2,F^3)$ and $F^0$ such that the systems with such
Hamiltonians be superintegrable.

Notice that symmetries  of systems with $2\times2$ and $3\times3$
matrix potentials  have been investigated in papers
\cite{nikitin:beckers1992} and \cite{nikitin:beckers1993}. However,
it was done only for diagonal potentials $\hat V$ depending on time
and one spatial variable.

\section{Determining equations}

Let us search for the first and second order integrals of motion for systems described by equation (\ref{SP}). By definition such integrals of motion $Q$ are first and second order differential operators commuting with Hamiltonian $H$:
\begin{gather}\label{com}[H,Q]\equiv HQ-QH=0.\end{gather}
We suppose these operators be formally self adjoint. Then, without loss of generality, they can be written in the following form:
\begin{gather}\label{qu}\begin{split}&
Q=\frac14\sigma^\mu\{\{\Phi^{\mu a b},\nabla_a \},\nabla_b \}+\ri\sigma^\mu \{\Lambda^{\mu a},\nabla_a\}+\sigma^\mu\Omega^\mu\end{split}\end{gather}
where $\Phi^{\mu a b}, \Lambda^{\mu a}$ and $\Omega^\mu$ are (unknown) real functions of $\bf x$, $\{\Phi^{\mu
a b},\nabla_a\}=\Phi^{\mu a b}\nabla_a+\nabla_a \Phi^{\mu a b},\
\nabla_a=\frac{\p}{\p x_a}$, $\sigma^\mu$ are Pauli matrices with $\sigma^0$ being the $2\times2$ unit matrix. In addition, here and in the following the summation is imposed over the repeated indices. Moreover, in all equations the Latin and Greek indices take the values 1,2,3 and 0,1,2,3 correspondingly.

Substituting (\ref{Redu}) and (\ref{qu}) into (\ref{com}),
using the relations
\begin{gather*}\sigma^a\sigma^b=\delta^{ab}+\ri\varepsilon^{abc}\sigma^c \end{gather*} where $\varepsilon^{m n k}$ is the Levi-Civita symbol,
and equating coefficients for linearly
independent matrices and differential operators, we obtain the
following system of determining equations for functions $\Phi^{\mu a b}$,
$\Lambda^{\mu a}$,  $\Omega^\mu$ and $F^\mu$:
\begin{gather}\la{e1}\Phi^{\mu a b}_c+\Phi^{\mu  b c}_a+ \Phi^{\mu c a}_b=0,
\\\la{e11}\Lambda^{0 a}_b+\Lambda^{0 b}_a=0,\\\la{e2}
\Lambda^{m a}_b+\Lambda^{m b}_a+\varepsilon^{m n k}B^k\Phi^{n a b}=0,\\\la{e3}\Phi^{0 a k}F^b_k+\Phi^{b a k}F^0_k-2\varepsilon^{b c k}\Lambda^{c a}F^k+\Omega^b_a=0,\\\la{e4}
\Lambda^{a k}F^0_k+\Lambda^{0k}F^a_k+\varepsilon^{a c k}\Omega^c F^k+\varepsilon^{n k a}\Phi^{n c d}F^k_{cd}=0,\\\la{e5}
\Phi^{\mu a k}F^\mu_k+\Omega^0_a=0,\\\la{e6}\Lambda^{\mu k}F^\mu_k=0.\end{gather}
Here the subindices denote
derivatives w.r.t. the  independent variables, i.e.,
$F^a_b=\frac {\p B^a}{\p x_b}$, etc.

Thus to classify Hamiltonians (\ref{Redu}) which admit first- and
second-order integrals of motion we are supposed to solve the system
of determining equations  (\ref{e1})-(\ref{e6}). Moreover, searching
for the first order integrals of motion we should a priori set
$\Phi^{\mu ab}=0$.

\section{Classification results}

The system (\ref{e1})-(\ref{e6}) is rather complicated and includes 77 coupled nonlinear partial differential equations for 44 variables. However, it is rather symmetric, and some of its constituents are easy integrable. The general solution of this system with two independent variables for the case $\Phi^{\mu a b}=0$ was found in \cite{N2}.

In this paper we find solutions of equations (\ref{e1})-(\ref{e6}) compatible with the supposition that Hamiltonian (\ref{Redu}) is invariant with respect to the rotation group O(3). In other words we suppose that $H$ commutes with generators of this group which are nothing but components of the total orbital momentum vector $\bf J$:
\begin{gather}\la{om} {\bf J}={\bf x}\times{\bf p}+\frac12\mbox{\boldmath $\sigma$}.\end{gather}

This condition reduces  the general form of external fields to:
\begin{gather}\la{F}F^0=\phi(x),\quad F^a=x^a\varphi(x)\end{gather}
where $\phi$ and $\varphi$ are functions of $x=\sqrt{x_1^2+x_2^2+x_3^2}$.

The system (\ref{e1})-(\ref{e6}) with additional conditions (\ref{F}) is algorithmically solvable, see Appendix.
 As a result we obtain the following list of Hamiltonians (\ref{Redu}) together with the admissible constants of motion additional to (\ref{om}):
 \begin{gather}\la{H1}\begin{split}&H=H_1=-\nabla^2+
\lambda\frac{\mbox{\boldmath$\sigma$}\cdot{\bf x}}{x^3}+\varphi(x),\\&
Q_1=\mbox{\boldmath$\sigma$}\cdot{\bf L}+1 +\lambda\mbox{\boldmath
$\sigma$}\cdot{\bf n};\end{split}\\\nonumber\\
\la{H2}\begin{split}&H=H_2=-\nabla^2+\mbox{\boldmath
$\sigma$}\cdot{\bf n}f'+f^2-\frac{\alpha}x,\\
&Q_2=\left(\ri{\mbox{\boldmath$\sigma$}}\cdot{\bf p}+f\right)\left({\mbox{\boldmath$\sigma$}}\cdot{\bf L}+1\right)+\frac\alpha2{\mbox{\boldmath$\sigma$}}\cdot{\bf n};\end{split}\\\nonumber\\
\la{H3}\begin{split}&H=H_3=-\nabla^2+\frac{\lambda}x\mbox{\boldmath
$\sigma$}\cdot{\bf n}+\frac{\lambda^2}{x^2}-\frac{\alpha}x,\\
&Q_1,\quad Q_3=Q_2|_{f=\frac\lambda{x}};\end{split}\end{gather}
 \begin{gather}\la{H4}\begin{split}&H=H_4=-\nabla^2+\lambda\frac{\mbox{\boldmath $\sigma$}\cdot {\bf x}}{x^2},\\&{\bf R}=\frac12({\bf p}\times{\bf J}-{\bf J}\times{\bf p})+\frac{{\lambda \bf x}\mbox{\boldmath $\sigma$}\cdot {\bf x}}
 {x^2}.\end{split}\end{gather}
 Here ${\bf L}={\bf x}\times{\bf p}$ is the orbital momentum,  ${\bf n}=\frac{\bf x}x$, $\varphi(x)$ and $f$ are arbitrary function of $x$, $f'=\frac{\p f}{\p x}$.

 Hamiltonian (\ref{H3}) is a particular case of operators presented in (\ref{H1}) and (\ref{H2}), which corresponds to a more extended number of integrals of motion.

 Hamiltonian (\ref{H4}) admits three integrals of motion, additional to (\ref{om}). They are components of vector $\bf R$, which generalizes the Laplace-Runge-Lenz vector to the case of a system with spin.

 The system with Hamiltonian (\ref{H4}) possesses the dynamical symmetry w.r.t. group O(4) whose generators are vectors $\bf J$  and $\bf R$. This system was discussed in paper \cite{N2} where its shape invariance was proven and exact solutions of the corresponding eigenvalue problem (\ref{SP}) where found.

 Notice that all Hamiltonians (\ref{H1})-(\ref{H4}) admit one more (discrete) symmetry. Namely, they commute with the following operator
\begin{gather}\la{P}Q_4=(\mbox{\boldmath $\sigma$}\cdot {\bf L}+1)p\end{gather}
where $p$ is the space inversion operator which changes the sign of independent variables, i.e., acts on the wave function as $p\psi({\bf x})=\psi(-{\bf x})$.

One more symmetry which includes the space reflection is valid for Hamiltonians $H_2$. Namely, these Hamiltonians commute with the following operator:
\begin{gather*}Q_5=(\mbox{\boldmath $\sigma$}\cdot {\bf p}+ f)p+\alpha\mbox{\boldmath $\sigma$}\cdot {\bf n}.\end{gather*}

Of course,  this symmetry with $f=\frac\lambda{x}$ is valid for Hamiltonian $H_3$.

\section{Algebraic properties of integrals of motion}
Let us discuss some general properties of the constants of motion presented in the previous section.

By construction, operators $Q_1,\ Q_2,\ { Q}_3$ and $\bf R$ commute with the corresponding Hamiltonians, i.e., satisfy conditions (\ref{com}).

Operators $Q_1,\ Q_2,$ and ${Q}_3$ are rotational scalars and so commute with the total orbital momentum (\ref{om}):
\begin{gather}\la{SCR}[Q_a,{\bf J}]=0, \quad a=1,2,3.\end{gather}

In addition, these operators satisfy the following algebraic relations:
\begin{gather}Q_1^2={\bf J}^2+\lambda^2+\frac14,\la{sq1}\\Q_2^2=\left({\bf J}^2+\frac14\right)H_2+\frac{\alpha^2}4, \la{sq2}\\Q_3^2=\left({\bf J}^2+\frac14\right)H_3+\frac{\alpha^2}4,\la{sq3} \\ Q_1Q_3+Q_3Q_1=\alpha\lambda,\quad [Q_1^2,Q_3]=0,\quad [Q_3^2,Q_1]=0.\la{sq4}\end{gather}

We see that the integrals of motion form rather non-trivial superalgebraic structures. Relations (\ref{sq3}) and (\ref{sq4}) will be used to explain the degeneration of spectrum of Hamiltonian (\ref{H3}).

In the particular case $\varphi=\frac\alpha{x^2}$  integrals of motion (\ref{om}),  (\ref{H1}) together with operators
  $D=x_1p_1+x_2p_2+x_3p_3$ and $K={x^2}/2$ form a basis of the seven-dimensional Lie
  algebra since the following  commutation relations are satisfied:
  \begin{gather}\begin{split}&  [H_1, D]=-2\ri H,\quad [K,H_1]=\ri
  D,\\&[J^a,J^b]=\ri\varepsilon^{abc}J^c,\quad [K,D]=2\ri K \end{split}\la{CA}\end{gather}
  while all the other commutators are trivial. In other words, there is a direct sum of the conformal algebra so(1,2)$\ni\langle H_1,\ D,\  K\rangle$, the Lie algebra of the rotation group so(3) $\ni \langle\ J^1,\ J^2,\ J^3\rangle$  and the one-dimension algebra spanned on $Q_1$. Thanks to the conformal symmetry   the discussed system can be interpreted as a model of conformal quantum mechanics, see \cite{burda} for definitions.

Finally, consider the superintegrable system (\ref{H4}) admitting vector integrals of motion.  Components ${ R}_a$ of vector operator ${\bf R}$ satisfy the following commutation relations:
\begin{gather}\la{core}\begin{split}&[J_a,J_b]=\ri\varepsilon_{abc}J_c,\quad
[ { R}_a,J_b]= \ri\varepsilon_{abc} { R}_c,\\&[ { R}_a, { R}_b]=-{2\ri}\varepsilon_{abc}J_c H_4.
\end{split}\end{gather}

Being considered on eigenvectors of Hamiltonian $H_4$  corresponding to coupled states, algebra (\ref{core}) is
isomorphic to the Lie algebra of group O(4). In other words, the system
with Hamiltonian $H_4$ admits the same dynamical symmetry as the Hydrogen atom. The detailed analysis of this system is presented in \cite{N2}.
\section{Exact solutions}

Since Hamiltonians $H_1$ and $H_2$ are defined up to arbitrary functions, they represent an infinite set of superintegrable models. In this section we consider two important particular cases of such models which involve the Coulomb potential, and present a constructive way  for finding solutions for the systems with arbitrary potentials.
\subsection{Charged particle with electric dipole moment interacting with  the field of point charge}
Let us start with Hamiltonian $H_1$. It includes an arbitrary scalar potential $\varphi$ and a dipole interaction term $\sim \mbox{\boldmath $\sigma$}\cdot {\bf F}$  with coupling constant $\lambda$ and external field ${\bf F}=\frac{\bf x}{x^3}$. This field can be interpreted as an electric field generated by a point charge. Thus it is naturally to choose $\varphi=\frac{\alpha}x$, then the corresponding operator $H_1$ can be interpreted as a Hamiltonian  of a charged  particle with spin 1/2 and a non-trivial
dipole electric moment.

Consider the eigenvalue problem for such specified Hamiltonian $H_1$:
\begin{gather}\left(-\Delta+\lambda\frac{\mbox{\boldmath
$\sigma$}\cdot{\bf x}}{x^3}-\frac{\alpha}x\right)\psi=\hat E\psi.\la{EP1}\end{gather}
Here $\psi=\psi({\bf x})$ is a two-component function which is supposed to be normalizable  and vanishing at $x=0$. In addition, to obtain a system with coupled states we suppose that $\alpha>0$.

Equation (\ref{EP1}) admits three  constants of motion $J_3,  {\bf J}^2$ and $Q_1$ which commute each other.
Thus we can  expand $\psi=\psi({\bf x})$ via eigenvectors $\hat\Omega_{j,\kappa,\nu}(\varphi,\theta)$ of these operators:
\begin{gather}\psi=\frac1x\sum_{j,\kappa,\nu} \psi_{j \kappa\nu}(x)
\hat\Omega_{j,\kappa,\nu}(\varphi,\theta).\la{es1}\end{gather} Here $x,\ \varphi$ and $\theta$ are spherical  coordinates,
$j=\frac12,\ \frac32,\ \dots, \ \kappa=-j,\ -j+1,\ \dots\ j$, and
$\nu=\varepsilon\mu$ (where $\mu=\sqrt{j(j+1)+\lambda^2+\frac14}$ and $\varepsilon=\pm1$) are quantum numbers which label the eigenvalues:
\begin{gather}\la{ev}\begin{split}&{\bf J}^2\hat\Omega_{j,\kappa,\nu}=
j(j+1)\hat\Omega_{j,\kappa,\nu},\\&J_3\hat\Omega_{j,\kappa,\nu}=
\kappa\hat\Omega_{j,\kappa,\nu},\end{split}\\\la{ev2}Q_1\hat\Omega_{j,\kappa,\nu}=
\nu\hat\Omega_{j,\kappa,\nu}.\end{gather}
The explicit form of $\hat\Omega_{j,\kappa,\nu}$ will be  specified later. Notice that the eigenvalues of $Q_1$ can be found algebraically starting with relation (\ref{sq1}) and the first of relations (\ref{ev}).

 Substituting
(\ref{es1}) into (\ref{EP1}) we obtain the following ordinary differential equations for radial functions $\psi_{j\kappa\nu}$:
\begin{gather}\left(-\frac{\p^2}{\p x^2}+\frac{\nu(\nu+1)-\lambda^2 }{x^2}-\frac\alpha{x}
\right)
\psi_{j\kappa\nu}=\hat E\psi_{j\kappa\nu}.\la{rep}\end{gather}
This equation is solved by the following functions:
\begin{gather}\la{GG}\psi_{j\kappa\nu}=C_{j\kappa\nu}x^{k_\nu+\frac12}
\exp\left(\sqrt{-\hat E} x\right){\cal F}\left(a_\nu,2k_\nu+1,2\sqrt{-\hat E}x\right)\end{gather} where $C_{j\kappa\nu}$ is an integration constant, $\cal F$ is the confluent hypergeometric function, and \begin{gather}\la{GG1}k_\nu=\sqrt{\nu(\nu+1)-\lambda^2},\quad a_\nu= k_\nu+\frac12-\frac{1}{\sqrt{-\hat E}}, \quad \nu=\pm\sqrt{j(j+1)+\lambda^2+\frac14}.\end{gather}

In order function (\ref{GG}) be bounded at infinity, the argument $a_\nu$ has to be a negative integer or zero, i.e.,
\begin{gather}\la{GG2}k_\nu+\frac12-\frac{1}{\sqrt{-\hat E}}=-n,\quad n=1, 2, ... \end{gather}
The corresponding eigenvalue $\hat E$ in (\ref{EP1}) and the energy value $E$ from (\ref{change}) are given by the
following equations:
\begin{gather}\la{ee1}\hat E=-\frac{\alpha^2}{4N^2} \quad \text{and}\quad E=-\frac{m\alpha^2}{2N^2}, \end{gather}where
\begin{gather}\label{ee2}N=\sqrt{\nu(\nu+1)-\lambda^2}+n+\frac{1}2,\quad n=0, 1, 2,\dots\end{gather}

The energy levels (\ref{ee1}) are degenerated w.r.t. quantum number $\kappa$ which are eigenvalues of the third component of the total orbital momentum. And there are no other degenerations.

Let us present the basic functions $\hat\Omega_{j,\kappa,\nu}=\hat\Omega_{j,\kappa,\nu}(\varphi,\theta)$, which are used in formula (\ref{es1}):
 \begin{gather}\la{ome1}\hat\Omega_{j,\kappa,\nu}=\frac1{2\sqrt{\mu}}
 \left(\sqrt{\frac{\left(\mu+j+\frac12\right)(j+\kappa)}j}Y_{j-\frac12,\kappa-\frac12}+
 \lambda\sqrt{{\frac{j-\kappa+1}{(j+1)\left(\mu+j+\frac12\right)}}}
 Y_{j+\frac12,\kappa-\frac12}\right)\end{gather}
if $ \nu=\mu>0$, and
 \begin{gather}\la{ome2}\hat\Omega_{j,\kappa,\nu}=\frac1{2\sqrt{\mu}}
 \left(\lambda\sqrt{\frac{(j-\kappa)}{j\left(\mu+j+\frac12\right)}}
 Y_{j-\frac12,\kappa+\frac12}+
 \sqrt{{\frac{(j+\kappa+1)\left(\mu+j+\frac12\right)}{(j+1)}}}
 Y_{j+\frac12,\kappa+\frac12}\right)\end{gather}if $\nu=-\mu<0$. Here $ Y_{j\pm \frac12,
k\pm \frac12}$ are spherical functions.
\subsection{Supersymmetric system}
Let us discuss the maximally superintegrable system whose Hamiltonian $H_3$ is defined by equation (\ref{H3}), and consider the related eigenvalue problem:
\begin{gather}H_3\psi\equiv\left(-\Delta+\frac{\lambda}x\mbox{\boldmath
$\sigma$}\cdot{\bf n}+\frac{\lambda^2}{x^2}-\frac{\alpha}x\right)\psi=\hat E\psi.\la{EP3}\end{gather}

Equation (\ref{EP3}) admits a number of symmetry operators given by relations (\ref{om}) and (\ref{H3}). Among them there are two commuting integrals of motion $J_3,\ {\bf J}^2$ and two anticommuting constants of motion $Q_1$ and $Q_3$. Symmetries $J_3$ and $ {\bf J}^2$ make it possible to separate variables. The constants of motion  $Q_1$ and $Q_3$ enable to decouple the system of equations in radial variables. In addition, it will be shown that these constants of motion generate a specific degeneration of the energy spectrum.

Like (\ref{EP1}),
equation (\ref{EP3}) is exactly solvable. Expanding its solutions via basis vectors (\ref{ome1}) and (\ref{ome2}), i.e., using representation   (\ref{es1}), we come to the following equations for radial functions:
\begin{gather}{\cal H}_\nu\psi_{j\kappa\nu}\equiv\left(-\frac{\p^2}{\p x^2}+\frac{\nu(\nu+1) }{x^2}-\frac\alpha{x}
\right)
\psi_{j\kappa\nu}=\hat E\psi_{j\kappa\nu}.\la{repka}\end{gather}
where $\nu$ are parameters defined in (\ref{GG1}).

Hamiltonian $\cal H_\nu$ is shape invariant. In other words, it can be factorized:
\begin{equation}\label{f}{\cal H}_{\nu}=a_\nu^+a_\nu+c_\nu\end{equation} where
\begin{gather}\label{a}a_\nu=\frac{\partial}{\partial x}+W_\nu,\  \
a_\nu^+=- \frac{\partial}{\partial x}+W_\nu,\ \
c_\nu=-\frac{\alpha^2}{\left(|\nu|+\frac{\varepsilon+1}2\right)^2},\ \varepsilon=\text{sign}~\nu,\end{gather} and  $W_\nu$ is a superpotential:
\begin{gather*}W_\nu=
\frac{2\alpha}{2|\nu|+\varepsilon+1}-\frac{2|\nu|+\varepsilon+1}{4x},\quad .\end{gather*} Moreover, operators  ${\hat{\cal
H}}_\nu={\cal H}_{\nu}-c_\nu$  satisfy the  intertwining
relations
\begin{gather}\la{ir}{\hat{\cal H}}_\nu a^+_\nu=a^+_\nu{\hat{\cal H}}_{\nu+1}.\end{gather}
This means that eigenvalues of Hamiltonian $\cal H_\nu$ (\ref{repka}) can be found algebraically using tools of supersymmetric quantum mechanics. As a result we obtain these eigenvalues in the form (\ref{ee1}) where
\begin{gather}\label{ee3}N=n+|\nu|+\frac{\varepsilon+1}2,\quad n=0, 1, 2,\dots\end{gather}

Like in the Hydrogen atom, energy values (\ref{ee1}), (\ref{ee3}) are proportional to the inverse square of the main quantum number $N$. But in contrast with the Hydrogen atom, $N$ is not a linear combination of two independent non-negative integers, and so there is no an $N$-fold degeneration. However, since $\nu$ can take positive and negative values as well, there is a two-fold degeneration typical for supersymmetric systems. In addition, since the quantum number $\kappa$ does not affect the energy values, we have the additional $(2j+1)$-fold degeneration w.r.t. $\kappa$.

Let us show that the supersymmetric nature of spectrum (\ref{ee1}), (\ref{ee3}) is caused integrals of motion $Q_1$ and $Q_3$.

It follows from (\ref{sq4}) that operator $Q_4=\frac\ri2[Q_1,Q_3]$ anticommutes with $Q_1$ and $Q_3$. On the set of eigenfunctions of operators ${\bf J}^2$ and $Q_1$ it is possible to define the  rescaled operators
\begin{gather}\la{rsc}\hat Q_3=\frac{Q_3}{j+\frac12},\quad \hat Q_4=\frac{Q_4}\nu+\frac{\alpha Q_1}{(2j+1)\nu}\end{gather} which satisfy the following anticommutation  relations
\begin{gather}\la{ar}\hat Q_a\hat Q_b+\hat Q_b\hat Q_a=2\delta_{ab}\hat H.\end{gather}
Here any of subindices $a$ and $b$ independently takes the values 3 and 4, and $\hat H=H_3+\frac{\alpha^2}{(2j+1)^2}$. By construction both $\hat Q_3$ and $\hat Q_4$ commute with $\hat H$, but this fact is also a consequence of (\ref{ar}).

Thus the rescaled integrals of motion $\hat Q_3$, $\hat Q_4$ and Hamiltonian $\hat H$ form a basis of the superalgebra of supersymmetric quantum mechanics. The two-fold  degeneration of spectrum of Hamiltonian  $\hat H$ (and so of Hamiltonian $H_3$) is a direct consequence of algebraic relations (\ref{ar}).

Notice that the ground state with energy $\hat E=\frac{\alpha^2}{(2\nu)^2}$ (see equations (\ref{ee1}), (\ref{ee3}) for $n=0$ and $\varepsilon=-1$) is not degenerated. Thus the supersymmetry of system (\ref{EP3}) is exact.

Let us also present the  state vectors corresponding to eigenvalues (\ref{ee1}), (\ref{ee2}):
 \begin{gather}\la{psin2}\psi^{(n)}_{j\kappa\nu}=C_nx^{|\nu|+\frac12}
 \exp\left(-\sqrt{-\hat E}x\right){\cal F}\left(-n,2|\nu|+1,\sqrt{-\hat E}x\right). \end{gather}
 Like in (\ref{GG}),  ${\cal F}(-n,2|\nu|+1,x)$ is the confluent hypergeometric function. However, its arguments differ from arguments of function (\ref{GG}).
 \subsection{Equations for radial wave functions for arbitrary potentials}
 Eigenvalue problem for Hamiltonians $H_1$ and $H_2$ can be effectively decoupled for the case of arbitrary function $\varphi$ and $f$ present in their definitions  (\ref{H1}) and (\ref{H2}). Here we deduce the decoupled equations for radial wave functions.

 For Hamiltonian $H_1$ it can be done in complete analogy with section 6.1. Considering the eigenvalue problem for this Hamiltonian with arbitrary function $\varphi$ and repeating all steps presented in this section before equation (\ref{rep}), we obtain the following equation:
 \begin{gather}{\cal H}_\nu\psi_{j\kappa\nu}\equiv\left(-\frac{\p^2}{\p x^2}+\frac{\nu(\nu+1)-\lambda^2}{x^2}+\varphi(x)
\right)
\psi_{j\kappa\nu}=\hat E\psi_{j\kappa\nu}\la{repa}\end{gather}
where $\nu$ is the parameter defined in (\ref{GG1}).

Thus expanding solutions via eigenvectors of the commuting integrals of motion, satisfying (\ref{ev}) and (\ref{ev2}), it is possible to reduce the eigenvalue problem for Hamiltonian $H_1$ (\ref{H1}), which is a three dimensional system of two coupled equations, to the infinite set of decoupled ordinary differential equations (\ref{repa}).

In order radial functions $\psi_{j\kappa\nu}$ to have a good behavior at $x=0$, the potential $\varphi(x)$ should increase not faster than $\frac{\alpha}{x^2}$ when $x\to0$. If for small $x$ this potential increases as $\frac{\alpha}{x^2}$, then parameter $\alpha$ should satisfy the condition $\alpha>\sqrt{1+\lambda^2}-\frac34$.

For some particular potentials $\varphi(x)$ equations (\ref{repa}) can be solved explicitly. Examples of such potentials are: \begin{gather}\la{ex1}\varphi(x)=-\frac{\alpha}x,\\\la{ex2} \varphi(x)=\frac{\alpha}x+\frac\beta{x^2},\quad \beta\neq0,\\\la{ex3} \varphi(x)=\omega^2x^2.\end{gather}

In the present paper only solutions corresponding to the Coulomb potential (\ref{ex1}) are discussed, see section 6.1.

Consider now the eigenvalue problem for Hamiltonian $H_2$ with arbitrary potential:
\begin{gather}\la{ep4}H_2\psi=\left(-\nabla^2+\mbox{\boldmath
$\sigma$}\cdot{\bf n}f'+f^2-\frac{\alpha}x\right)\psi=\hat E\psi.\end{gather}
In addition to (\ref{ep4}), we impose on wave function $\psi$ the following condition:
 \begin{gather}\la{ev3}Q_2\psi=\left(\left(\ri{\mbox{\boldmath$\sigma$}}\cdot{\bf p}+f\right)\left({\mbox{\boldmath$\sigma$}}\cdot{\bf L}+1\right)+\frac\alpha2{\mbox{\boldmath$\sigma$}}\cdot{\bf n}\right)\psi=q\psi\end{gather}
where eigenvalues $\hat E$ and $q$ are connected by the following algebraic relation:
\begin{gather}\la{E_nu}4q^2=\left(2j+1\right)^2\hat E+\alpha^2.\end{gather}

Equations (\ref{ep4}) and (\ref{ev3}) are compatible since operators $H_2$ and $Q_2$ commute each other and satisfy relations (\ref{H2}).

To separate variables, let us expand solutions of equations (\ref{ep4}) and (\ref{ev3}) via the complete
 set of eigenfunctions of the commuting operators $ {\bf J}^2,\ J_3,$ and $\mbox{\boldmath
$\sigma$}\cdot{\bf L}+1 $:
 \begin{gather}\psi=\frac1r\sum_{j,\kappa,\varepsilon\mu} \psi_{j, \kappa,\varepsilon\mu}(r)
\Omega_{j,\kappa,\varepsilon\mu}(\varphi,\theta).\la{SOS}\end{gather} Here $\Omega_{j,\kappa,\varepsilon\mu}(\varphi,\theta)$ are spherical spinors,
$j, \ \kappa,\ \varepsilon=\pm1$ and  $ \mu=j+\frac12$ are quantum numbers labeling eigenvalues of above mentioned operators:
\begin{gather*}\begin{split}&{\bf J}^2\Omega_{j,\kappa,\varepsilon\mu}=
j(j+1)\Omega_{j,\kappa,\varepsilon\mu},\\&J_3\Omega_{j,\kappa,\varepsilon\mu}=
\kappa\Omega_{j,\kappa,\varepsilon\mu},\\&\left(\mbox{\boldmath
$\sigma$}\cdot{\bf L}+1\right)\Omega_{j,\kappa,\varepsilon\mu}=
\varepsilon\mu\Omega_{j,\kappa,\varepsilon\mu}
\end{split}\end{gather*}

  As a result equation (\ref{ep4}) is reduced to the following coupled  system of ordinary differential equations for the radial wave function:
  \begin{gather}\la{ep5}\left(-\frac{\p^2}{\p x^2}+\frac{\mu(\mu+1)}{x^2}+f^2-\frac\alpha{x}\right)\psi_{j, \kappa,\mu}(x)+f'\psi_{j,\kappa,-\mu}(x)=\hat E\psi_{ j, \kappa,\mu}(x),\\\left(-\frac{\p^2}{\p x^2}+\frac{\mu(\mu-1)}{x^2}+f^2-\frac\alpha{x}\right)
  \psi_{j,\kappa,-\mu}(x)+f'\psi_{j ,\kappa,\mu}(x)=\hat E\psi_{ j, \kappa,-\mu}(x).\la{ep6}\end{gather}

  Using (\ref{ev3}) this system can be decoupled.
Substituting (\ref{SOS}) into (\ref{ev3}) and turning into account the relations
\begin{gather*}{\mbox{\boldmath$\sigma$}}\cdot{\bf p}={\mbox{\boldmath$\sigma$}}\cdot{\bf n}{\mbox{\boldmath$\sigma$}}\cdot{\bf n}{\mbox{\boldmath$\sigma$}}\cdot{\bf p}=\left({\bf x}\cdot{\bf p}-\frac{\ri}x\right){\mbox{\boldmath$\sigma$}}\cdot{\bf n}+
\frac{\ri}x{\mbox{\boldmath$\sigma$}}\cdot{\bf n}\left({\mbox{\boldmath$\sigma$}}\cdot{\bf L}+1\right),\\{\mbox{\boldmath$\sigma$}}\cdot{\bf n}\left({\mbox{\boldmath$\sigma$}}\cdot{\bf L}+1\right)=-\left({\mbox{\boldmath$\sigma$}}\cdot{\bf L}+1\right){\mbox{\boldmath$\sigma$}}\cdot{\bf n},\\\left({\mbox{\boldmath$\sigma$}}\cdot{\bf L}+1\right)^2={\bf J}^2+\frac14,\quad ({\mbox{\boldmath$\sigma$}}\cdot{\bf n})^2=1,\\
{\mbox{\boldmath$\sigma$}}\cdot{\bf n}\Omega_{j,\kappa,\varepsilon\mu}=\Omega_{j,\kappa,-\varepsilon\mu},
\quad
{\bf x}\cdot{\bf p}\Omega_{j,\kappa,\varepsilon\mu}=0,\quad {\bf x}\cdot{\bf p}\psi_{ j,\kappa,\varepsilon\mu}=-\ri\frac{\p}{\p x}\psi_{ j,\kappa,\varepsilon\mu}\end{gather*}
we obtain the following system of first order equations:
\begin{gather}\la{ev8}(f-\tilde q)\psi_{ j,\kappa,\mu}+a_\mu^+\psi_{ j,\kappa,-\mu}=0,\\\la{ev9}-(f+\tilde q)\psi_{ j,\kappa,-\mu}+a_j\psi_{ j,\kappa,\mu}=0\end{gather}
were
\begin{gather*}a_\mu=\left(\frac{\p}{\p x}+\frac{\mu}{x}-\frac{\alpha}{2\mu}\right),\qquad a^+_\mu=\left(-\frac{\p}{\p x}+\frac{\mu}{x}-\frac{\alpha}{2\mu}\right)\end{gather*}
and
\begin{gather*}\tilde q=\frac{2q}{2j+1}=\pm\left(\hat E+\frac{\alpha^2}{(2j+1)^2}\right)^\frac12.\end{gather*}

Solving equation (\ref{ev9}) for $\psi_{j,\kappa,-\mu}$ and substituting the derived expression into (\ref{ev8}) or into (\ref{ep5}) we obtain a decoupled system of second order equations:
\begin{gather}\la{eqq}\left(a^+_\mu a_\mu+\frac{f'}{f+\tilde q}a_\mu+f^2-\frac{\alpha^2}{4\mu^2}\right)\psi_{ j,\kappa,\mu}=\hat E \psi_{j,\kappa,\mu}.\end{gather}

Thus equation (\ref{ep4}) with arbitrary potential function $f=f(x)$ can be reduced to the decoupled system of ordinary differential equations (\ref{eqq}) for radial functions. However, it is doubtful whether the latter equations can be solved exactly for a fixed function $f=f(x)$ if $\alpha ff'\neq0$.

To end this section, let us write system (\ref{ep5}), (\ref{ep6}) as a single  equation with a matrix potential:
\begin{gather}\la{last}{\cal H}\tilde\psi_{j,\kappa,\mu}\equiv\left(-\frac{\p^2}{\p x^2}+U\right)\tilde\psi_{j,\kappa,\mu}=
\left(\hat E+\frac{\alpha^2}{(2j+1)^2}\right)
\tilde\psi_{j,\kappa,\mu}\end{gather}
where $\tilde\psi_{j,\kappa,\mu}=\text{column}
(\psi_{j,\kappa,\mu},\psi_{j,\kappa,-\mu})$, and \begin{gather*} U=W^2-W',\quad W=\left(\frac{\mu}{x}-\frac{\alpha}{2\mu}\right)\sigma_3+ f\sigma_1.\end{gather*}
In other words, the effective potential $U$ can be expressed via superpotential $W$. This circumstance makes it possible to find the  ground states $\psi^{(0)}_{j,\kappa,\mu}$ of Hamiltonians $\cal H$ which are solutions of equation $\left(\frac{\p}{\p x}+W\right)\tilde\psi^{(0)}_{j,\kappa,\mu}=0$. The latter equation is nothing but the system (\ref{ev8}), (\ref{ev9}) with $\tilde q=0$.
\section{Discussion}

We present the completed list of 3d Hamiltonians (\ref{Redu}) with matrix potentials, which are invariant w.r.t. the rotation group and admit first and second order integrals of motion. This list is given by equations (\ref{H1})-(\ref{H4}) and includes four representatives, two of which are defined up to arbitrary functions.

The presented systems have a clear physical interpretation and describe particles with spin 1/2 and non-trivial dipole moment. The system (\ref{H4}) was discussed in paper \cite{N2}. Like the Hydrogen atom,  it admits six integrals of motion satisfying algebra o(4). However, in contrast with symmetries of the Hydrogen atom, the Laplace-Runge-Lenz vector ${\bf R}$ in (\ref{H4}) is dependent on spin. In the present paper we prove that this system is unique, and there are no other 2$\times$2 matrix potentials compatible with this symmetry.

It is possible to verify by direct calculations that Hamiltonians $H_1..H_4$ do commute with the presented integrals of motion and do not admit another integrals of motion  independent with ones given by relations (\ref{H1})-(\ref{H4}). A much more difficult problem  is to prove that the list of superintegrable systems (\ref{H1})-(\ref{H4}) is complete. To do this it is necessary to find all non-equivalent solutions of the determining equations (\ref{e1})-(\ref{e6}). The related detailed calculations can be found in the Appendix.

Solving the determining equations we did not restrict ourselves to scalar and vector integrals of motion. In contrary, we consider also the tensor integrals of motion and prove, that for equation (\ref{Redu}) they do not exist.

The eigenvalue problems for Hamiltonians (\ref{Redu}) are systems of coupled partial differential equations of second order,  which are rather complicated. However, these Hamiltonians admit at least three commuting integrals of motion. Thank to this fact the eigenvalue problems can be effectively decoupled and reduced to ordinary differential equations in radial variables. Solutions for two of  the eigenvalue problems which include the Coulomb potential are presented in section 6, while the eigenvalues and eigenvectors of Hamiltonian (\ref{H4}) have been found in paper \cite{N2}.

In contrast with the case of scalar classical and quantum superintegrable systems, the systems with spin have been classified only partially. Thus they belong to  a perspective research  field, and it would be interesting to extend our knowledge of these interesting and important subjects. In particular, it is desirable to classify second order superintegrable systems  with dipole (and spin-orbit) interactions, which are not a priori rotationally invariant. This problem is much more complicated than in the case of scalar systems, and it is a challenge to solve it.

One more interesting task is to search for relativistic counterparts
of non-relativistic superintegrable  systems. The relativistic
analogues of the Pron'ko-Stroganov system \cite{Pron} and the system
with Hamiltonian (\ref{H4}) are  discussed in papers \cite{N10} and
\cite{N2}, and it is not too difficult to extend this discussion to the cases presented in (\ref{H1})-(\ref{H3}).

\appendix
 \section{Solution of the determining equation}
 Here we present the main steps in solution of the determining equations (\ref{e1})-(\ref{e6}).
 \renewcommand{\theequation}{A\arabic{equation}} \setcounter{equation}{0}

 \subsection{Decoupling of the determining equations}
First we will show how the rather complicated system of determining equations (\ref{e1})-(\ref{e6}) can be simplified and decoupled to algorithmically solvable subsystems.

 Let us start with the first order integrals of motion. The corresponding functions $\Phi^{\mu ab}$ are equal to zero, and equations (\ref{e1})-(\ref{e2}) are reduced to the following ones:
\begin{gather}\la{La}\Lambda^{\mu a}_b+\Lambda^{\mu b}_a=0.\end{gather}
These equations are easily integrated:
\begin{gather}\la{Kv}\Lambda^{0a}=\varepsilon^{abc}x_c\alpha^{ b}+
\nu^{a},\quad\Lambda^{ma}=\sum_i^6\Lambda^{ma}_i\end{gather}
where
\begin{gather}\la{KvvK1}\Lambda_1^{ma}=
\nu\delta^{m a},\quad \Lambda_2^{ma}=\mu\varepsilon^{m ac}x^c,\\\la{KvvK2} \Lambda_3^{ma}= \varepsilon^{m ac}\mu^c,\quad \Lambda_4^{ma}= \delta^{ma}\nu^cx_c-x^m\nu^a,\\\la{KvvK3} \Lambda_5^{ma}=\varepsilon^{abc}x_b\mu^{mc},\quad \Lambda_6^{ma}=\nu^{ma}.\end{gather}
Here $\alpha^b,\ \nu^a,\ \nu,\ \mu,\  \mu^c,\ \nu^{ma}$ and $\mu^{ma}$ are integration constants. Moreover, without loss of generality we can set $\alpha^b=0$ since non-trivial $\alpha^b$ correspond to already declared integrals of motion (\ref{om}).

Thus to find the first order integrals of motion it is sufficient to solve equations (\ref{e3})-(\ref{e6}) with the given coefficients (\ref{Kv}) and trivial $\Phi^{\mu ab}$. Moreover, all cases enumerated in (\ref{KvvK1}), (\ref{KvvK2}) and (\ref{KvvK3}) (which correspond to scalar, vector and tensors integrals of motion) should be considered separately.

Consider the second order integrals of motion. In accordance with
(\ref{e1}) functions $\Phi^{0 a b}$ should satisfy equations for
Killing tensors of rank 2. Thus  they are second order polynomials
in $x^a$ which can be represented in the following form \cite{NP}
\begin{gather}\la{Kt0}\Phi^{0 a b}=\Phi_1^{0 a b}+\Phi_2^{0 a b}+\Phi_3^{0 a b}+\Phi_4^{0 a b}\end{gather}
where
\begin{gather}\la{Kt1}\Phi_1^{0 a b}=\lambda_1\delta^{ab}+
\lambda_2(\delta^{ab}x^2-x^{a}x^{b}),\qquad \text{(scalar, even)} \\\la{Kt2}
\Phi_2^{0 a b}=\lambda_0^{a}x^b+
\lambda_0^{b}x^a-2\delta^{ab}\lambda_0^{c}x_c,\ \
\qquad \text{(vector, odd)} \\
\la{Kt3}\begin{split}&\Phi_3^{0 a b}=\lambda_1^{ ab}+\lambda^{ab}_2x^2-\lambda_2^{ac}x^bx_{c}\\&-\lambda_2^{bc}x^ax_{c}+
\delta^{ab}\lambda^{cd}_2x_{c} x_{d},\end{split}\quad\ \ \qquad \text{(tensor, even)}\\\la{Kt4}\Phi_4^{0 a b}=\lambda^{ ac}_3\varepsilon^{cbd} x_d+\lambda_3^{ bc}\varepsilon^{cad} x_d.\qquad \qquad \text{(tensor, odd)}\end{gather}
Here $\lambda_1, \lambda_2, \lambda^c, \lambda_c^{ab}$ are (real) integration constants. Moreover, $\lambda_c^{ab}$ are symmetric and traceless tensors. In the right brackets the covariant properties of integration constants and parities of $\Phi_i^{0 a b}$ as functions of $x_a$ are indicated.

Functions $\Phi^{m a b}$ with a fixed value of $m\neq0$ also are Killing tensors of rank 2 w.r.t. indices $a$ and $b$. Their general form is analogous to (\ref{Kt0}) but more complicated thanks to the additional    free index $m$:
\begin{gather}\la{Kt}\Phi^{m a b}=\sum_{i=1}^7\Phi_i^{m a b}\end{gather}
where
\begin{gather}\la{Ktt1}\Phi_1^{m a b}=\lambda(2x^m\delta^{ab}-x^a\delta^{m b}-x^b\delta^{m a}),\phantom{x^a\delta^{m b}-x^b\delta^{m a}aaaaaaaaaaa} \text{(scalar, odd)} \\
\la{Ktt2}\begin{split}
&\Phi_2^{m a b}=\lambda_1^m\delta^{ab}+\delta^{ma}\lambda_2^b+\delta^{mb}\lambda_2^a+
\lambda_3^m(\delta^{ab}x^2-x^{a}x^{b})\\&\phantom{ n=2}+
\delta^{ma}(x^b\lambda_4^cx_c-\lambda_4^bx^2)+
\delta^{mb}(x^a\lambda_4^cx_c-\lambda_4^ax^2)\\&\phantom{ n=2}-x^m(2\delta^{ab}\lambda_4^cx_c
-\lambda_4^ax^b-\lambda_4^bx^a),\end{split}\phantom{x^a\delta^{m b}-x^b\delta a}\text{(vector, even)}\\\la{Ktt3}\Phi_3^{m a b}=(\varepsilon^{mca}\lambda_5^b+
\varepsilon^{mcb}\lambda_5^a)x_c+\lambda_6^k(\delta^{ma}\varepsilon^{bck}+
\delta^{mb}\varepsilon^{ack})x_c,\phantom{x^a\delta^{m b}aa a}\text{(vector, odd)}\\\la{Ktt4}\begin{split}&\Phi_4^{m a b}=\varepsilon^{mac}\lambda_3^{cb}+\varepsilon^{mbc}\lambda_3^{ca}
+(\delta^{ma}\varepsilon^{dcb}+
\delta^{mb}\varepsilon^{dca})x_c\lambda_4^{dk}x_k\\&\phantom{ n=2}-
\lambda_4^{ad}x^m\varepsilon^{bdc}x_c-
\lambda_4^{bd}x^m\varepsilon^{adc}x_c,\end{split}
\phantom{x^aaad}\text{(rank 2 tensor, even)}\\\la{Ktt5}\begin{split}&
\Phi_5^{m a b}=
\lambda_1^{m a}x^b+\lambda_1^{m b}x^a-2\delta^{ab}\lambda_1^{m c}x_c\\&\phantom{ n=2}+2x^m\lambda_2^{ab}-
(\delta^{ma}\lambda_2^{bc}+\delta^{mb}\lambda_2^{ac})x_c,\end{split}
\phantom{x^aaaaaaaaaaaaaaaaa}\text{(rank 2 tensor, odd)}
\\\la{Ktt6}
\Phi_6^{m a b}=\lambda_2^{m ac}\varepsilon^{cbd} x_d+\lambda_2^{m bc}\varepsilon^{cad} x_d,\phantom{x^aaaaaaaaaaaaaaaaaaaaaa}\text{(rank 3 tensor, odd)}\\\la{Ktt7}\Phi_7^{m a b}=\lambda^{m ab}_3x^2-\lambda_3^{m ac}x^bx_{c}-\lambda_3^{m bc}x^ax_{c}
+\delta^{ab}\lambda^{m cd}_3x_{c} x_{d}.\phantom{x^aaaaa}\text{(rank 3 tensor, even)}
\end{gather}

Formulae (\ref{Kt0}) and (\ref{Kt}) give the general solution of equations (\ref{e1}). One more subsystem of the determining equations which can be easy integrated is given by formula (\ref{e11}). In this case we deal with the equation for  Killing vectors,  whose general solution is given by equation (\ref{Kv}).

Solving of other equations (\ref{e3})-(\ref{e6}) is a much more complicated problem which, however, can be effectively separated to relative simple subproblems starting with the following speculations.
\begin{itemize}
\item
By definition, Hamiltonian (\ref{Redu}) admits integrals of motion
(\ref{om}) and  is an integral of motion by itself. Thus without
loss of generality we can set $ \alpha^b=0$, $\Phi_1^{0 a b}=0$ and
$\lambda_2^{ab}$  in (\ref{Kv}), (\ref{Kt0}) and (\ref{Kt3})
respectively, since they correspond to higher order terms of
operators $H,$ $J_a$ and $J_aJ_b$.
\item
 Functions  (\ref{Ktt1}) and (\ref{Kt2}), (\ref{Ktt2}), (\ref{Ktt3})  correspond to scalar and vector integrals of motion while the remaining solutions  generate tensor operators (\ref{qu}). Since scalars, vectors and tensors transform in different way under rotation transformations which keep Hamiltonian invariant, all of them should satisfy the commutativity condition (\ref{com}) independently. In other words, the determining equations (\ref{e1})-(\ref{e6}) should be solved separately for scalar, vector and tensor operators.
\item Integrals of motion which are  tensors of rank 3 are forbidden since the number  of their components exceeds the maximal admissible number of integrals of motion. Thus it is possible a priori to set  in (\ref{Kt}) $\Phi_6^{m a b}=\Phi_7^{m a b}=0.$

\item\la{4} In accordance with (\ref{F}) the external field $F^a$ and potential $F^0$ are vector and scalar with well defined parities. Parities of solutions (\ref{Kt2})-(\ref{Kt4}) and (\ref{Ktt1})-(\ref{Ktt7}) are transparent also. Then, analyzing properties of  the remaining equations (\ref{e2})-(\ref{e6}) under the space inversion we conclude that  functions $\Phi^{mab}$, $\Lambda^{ma}$ and $\Omega^m$ should have the same parity. In addition, the parity of functions $\Phi^{0ab}$, $\Lambda^{0a}$ and $\Omega^0$ should be opposite to the parity of $\Phi^{mab}$, $\Lambda^{ma}$ and $\Omega^m$.  Thus the system (\ref{e1})-(\ref{e6}) should be solved separately for the cases when ($\Phi^{0ab}$, $\Lambda^{0a}$, $\Omega^0$) are even  and odd with ($\Phi^{mab}$, $\Lambda^{ma}$, $\Omega^m$) being odd and even respectively.
\end{itemize}
In accordance with the above the system of determining equations (\ref{e2})-(\ref{e6}) is decoupled to five subsystems corresponding to scalar, vector and tensor functions  $\Phi^{\mu ab}$  with fixed parities. In other words we are supposed to solve these equations with given combinations of functions $\Phi^{0ab}$, $\Phi^{mab}$ and $\Lambda^{0a}$, namely:
\begin{gather}\la{v11}\Phi^{0ab}=0,\qquad \Phi^{mab}=\Phi_1^{mab},
\\\la{v1}\Phi^{0ab}=\Phi_2^{0ab},\quad \Phi^{mab}=\Phi_2^{mab},\\\la{v2}
\Phi^{0ab}=0,\qquad \Phi^{mab}=\Phi_3^{mab},\quad \Lambda^{0a}=
\lambda^{a},\\\la{v3}
\Phi^{0ab}=\Phi_3^{0ab},\quad \Phi^{mab}=\Phi_5^{mab},\\\la{v4} \Phi^{0ab}=\Phi_4^{0ab},\quad \Phi^{mab}=\Phi_4^{mab}\end{gather}
where functions $\Phi_i^{\mu ab}$ are defined by equations (\ref{Kt2})-(\ref{Kt4}) and (\ref{Ktt1})-(\ref{Ktt5}).
Except the case (\ref{v2}) the  tensor $\Lambda^{0a}$
should be trivial.

\subsection{Scalar constants of motion}

Consider first order scalar constants of motion.  The corresponding coefficient functions  $\Lambda^{\mu a}$ are reduced to tensors $\Lambda_1^{m a}$ and $\Lambda_2^{m a}$ given in (\ref{KvvK1}). The general form of scalar operator (\ref{qu}) with such coefficient functions is:
\begin{gather}\la{quq}
Q=\nu_1\mbox{\boldmath
$\sigma$}\cdot{\bf L} +\nu_2\mbox{\boldmath
$\sigma$}\cdot{\bf p}+\mbox{\boldmath
$\sigma$}\cdot{\bf x}f^1(x)+f^2(x)\end{gather} where $f^1$ and $f^2$ are arbitrary functions of $x$. The corresponding coefficient functions in (\ref{qu}) are:
\begin{gather}\la{cf}\Lambda^{ma}=
\nu_1\delta^{m a}+\nu_2 \varepsilon^{m ac}x_c,\quad\Omega^0=f^2, \quad \Omega^m=x^mf^1 \end{gather}
while functions $\Phi^{\mu ab}$ are trivial.

Substituting (\ref{quq}), (\ref{cf}) into equations (\ref{e5}) and (\ref{e6}) we obtain that $f^2=$Const and $\nu_2=0$. Then equation (\ref{e4}) turns to identity, and the last remaining equation (\ref{e3}) is solved by $f_1=\frac\alpha{x}$ and $\varphi_1=\frac\alpha{x^3}$. As a result we obtain integrals of motion (\ref{H1}).

Consider second order scalars. There exist the only  second order scalar term  specified by the coefficient function (\ref{Ktt1}). The general form of the corresponding scalar operator is given by the following formula: \begin{gather}\la{qu22}
Q=\frac12(\mbox{\boldmath
$\sigma$}\cdot{\bf p}\times{\bf L}-\mbox{\boldmath
$\sigma$}\cdot{\bf L}\times{\bf p})+f(x)\mbox{\boldmath
$\sigma$}\cdot{\bf L} +\varphi(x)\mbox{\boldmath
$\sigma$}\cdot{\bf p}+g(x)\mbox{\boldmath
$\sigma$}\cdot{\bf x}+h(x)\end{gather}
where $f(x), \varphi(x), g(x)$ and $h(x)$ are functions of $x$ which will be specified in the following.

 The related multipliers $\Lambda^{ma}$, $\Omega^a$ and $\Omega^0$ in (\ref{qu}) are:
\begin{gather}\la{cf3}\Lambda^{ma}=
\varphi(x)\delta^{m a}+f(x) \varepsilon^{m ac}x_c,\quad\Omega^0=h(x), \quad \Omega^m=x^mg(x). \end{gather}

Substituting (\ref{F}), (\ref{Ktt1}) and (\ref{cf3}) into (\ref{e2}) we obtain the following conditions:
\begin{gather}\varphi(x)=0,\quad \la{phi}f(x)=x\varphi_1.\end{gather}

The next equation, i.e., (\ref{e3}), is solved by the following functions:
\begin{gather}\la{F0}g(x)=\frac\alpha{x},\quad F^0=f^2(x)-\frac\alpha{x}.\end{gather}

Equations (\ref{e4}) and (\ref{e6}) turn to identities, while the remaining equation (\ref{e5}) generates the following condition for function $h(x)$:
\begin{gather}\la{g3} h'(x)=2x\varphi_1=2f'(x).\end{gather}

Relations (\ref{cf3})-(\ref{g3}) together with (\ref{Redu}), (\ref{qu}) specify infinite many Hamiltonians defined up to arbitrary function $f$. The corresponding integrals of motion are given by equations (\ref{H2}) and (\ref{H3}).
\subsection{Vector integrals of motion}
\subsubsection{First order integrals}
We start with   first order vectors constants of motion. In accordance with (\ref{Kv}) and (\ref{KvvK2}) there are two possible sets of the corresponding non-trivial coefficient functions in (\ref{qu}):
\begin{gather}\la{cf2}\Lambda^{ma}= \varepsilon^{m ac}\mu^c,\quad \Omega^0=x_c\mu^cf_1,\ \quad \Omega^a=\mu^af_2+x^a\mu_cx^cf_3\end{gather}and
\begin{gather}\la{cf31}\Lambda^{0a}=\alpha\nu^{a},\quad \Lambda^{ma}= \delta^{ma}\nu^cx_c-x^m\nu^a,\quad \Omega^a=\varepsilon^{abc}x^b\nu^cf_4\end{gather}
where $f_1,...,f_4$ are functions of $x$. Functions $\Omega^a$ and $\Omega^0$ given in equations (\ref{cf2}) and (\ref{cf31}) represent generic scalars linear in vector parameters $\nu$ and $\mu$. Moreover, parities of these functions satisfy conditions requested in the last item in section A1.

Substituting (\ref{cf2}), (\ref{cf31}) and (\ref{F}) into determining equation (\ref{e3}) we obtain the necessary condition $\varphi=0$ which correspond to the trivial external field $\bf F$. Thus Hamiltonian (\ref{Redu}) with a nontrivial dipole interaction does not admit first order vector integrals of motion.
\subsubsection{General analysis of second order case}
Consider the second order vector constants of motion. The corresponding second order terms in (\ref{qu}) can include either even coefficients (\ref{v1}) or odd coefficients (\ref{v2}).

Let us start with even functions (\ref{v1}). Being included into the same integral of motion, all vector coefficients $\lambda_0^a,...,\lambda_4^a$ should be proportional, i.e., $\lambda_\mu^a=n_\mu\lambda^a$ with some constants $n_\mu$ and $\lambda^a$. Thus we have the following coefficient functions for second order terms:
\begin{gather}\la{cf5}\begin{split}&\Phi^{0 a b}=\nu_0(\lambda^{a}x^b+
\lambda^{b}x^a-2\delta^{ab}\lambda^{c}x_c),\\&\Phi^{m a b}=\nu_1\lambda^m\delta^{ab}+
\nu_2(\delta^{ma}\lambda^b+\delta^{mb}\lambda^a)+\nu_3(
\lambda^m(\delta^{ab}x^2-x^{a}x^{b}))\\&+\nu_4(
\delta^{ma}(x^b\lambda^cx_c-\lambda^bx^2)+
\delta^{mb}(x^a\lambda^cx_c-\lambda^ax^2) -x^m(2\delta^{ab}\lambda^cx_c
-\lambda^ax^b-\lambda^bx^a)).\end{split}\end{gather}

 The corresponding first order terms include even tensor functions $\Lambda^{ma}$ and odd vector functions $\Lambda^{0a}$. Moreover, all these functions should be linear in vector parameter $\lambda^a$, and $\lambda^{0a}$ should solve equation (\ref{e11}). Without loss of generality we can represent them in the following form: \begin{gather}\la{cf40}\Lambda^{0a}=\nu_5
\varepsilon^{abc}x^b\lambda^c,\\\la{cf4}\Lambda^{ma}=x^m\varepsilon^{ack}
\lambda^cx^kf_1+\varepsilon^{mka}x^k\lambda^cx^cf_2+
\varepsilon^{mab}\lambda^bf_3\end{gather} where $f_1, f_2$ and $f_3$
are arbitrary functions of $x$. Moreover, it is possible to set
$\nu_5=0$, since a nontrivial $\nu_5$ corresponds to derivative
terms of the total orbital momentum which is an integral of motion
by construction.

Vector function $\Omega^a$ and scalar $\Omega^0$ in (\ref{qu}) should be even and odd respectively. Their generic form is given by the following equation:
\begin{gather}\la{cf7} \Omega^0=\lambda^cx^cf_4,\quad \Omega^a=x^a\lambda^cx^cf_5+\lambda^af_6.
\end{gather}

Considering the odd coefficients (\ref{v2}) we have the following coefficient functions for second, first and zero order terms:
\begin{gather}\la{n1}\Phi^{m a b}=\nu_1(\varepsilon^{mca}\lambda^b+
\varepsilon^{mcb}\lambda^a)x_c+\nu_2\lambda^k(\delta^{ma}\varepsilon^{bck}+
\delta^{mb}\varepsilon^{ack})x_c,\\\la{n2}\Lambda^{ma}=
x^mx^a\lambda^cx^cf_1 +\delta^{ma}\lambda^cx^cf_2+x_m\lambda^af_3+x_a\lambda^mf_4,\\
\la{n3}\Lambda^{0a}=\nu_3\lambda^a,\quad
\Omega^a=\varepsilon^{abc}x^b\lambda^cf_5
\end{gather}
where $f_1..f_5$ are arbitrary functions of $x$.
The other  coefficients in (\ref{qu}) are trivial.

Thus the classification of second order vector integrals of motion is reduced to finding non-equivalent solutions of equations (\ref{e1})-(\ref{e6}) for functions (\ref{cf5})-(\ref{cf7}) and (\ref{n1})-(\ref{n3}).
\subsubsection{Solutions for functions (\ref{cf5})-(\ref{cf7}) with trivial $\Phi^{mab}$}
For the case of trivial $\Phi^{mab}$ we have the following coefficient functions allowed  by relations (\ref{e11}) and (\ref{e2}):
\begin{gather}\begin{split}\la{cf17}&\Phi^{0 a b}=\lambda^{a}x^b+
\lambda^{b}x^a-2\delta^{ab}\lambda^{c}x_c,\quad \Lambda^{ma}=
\nu\varepsilon^{mac}
\lambda^c\end{split}\end{gather}
while the generic form of $\Omega^a$ and $\Omega^0$ are still defined by equation (\ref{cf7}).

Substituting (\ref{F}), (\ref{cf17}) and (\ref{cf7}) into equation (\ref{e3}) we obtain:
\begin{gather*}\left(\frac12(x^a\lambda^b+x^b\lambda^a)-
\delta^{ab}x^c\lambda^c\right)\varphi+\frac12\left(x^2\nu^ax^b-x^ax^b
\lambda^cn^c\right)\frac{\varphi'}x\\-2\nu(x^a\lambda^b-\delta^{ab}\lambda^cx^c)
\varphi+\lambda^bx^a\frac{f_6'}x+\nu^ax^bf_5+\delta^{ab}x^c\lambda^cf_5+
x^ax^b\lambda^cx^c\frac{f_5'}x=0.\end{gather*} Equating coefficients
for linearly independent terms $x^a\lambda^b,\ x^b\lambda^a,\
x^ax^b\lambda^cx^c$ and $\delta^{ab}\lambda^cx^c$, we obtain the
following system of equations:
\begin{gather}\begin{split}&x\varphi(1-4\nu)+2f_6'=0,\quad (x\varphi)'+2f_5=0,\\&\varphi'=2f_5',\quad (2\nu-1)\varphi+f_5=0.\end{split}\la{cf18}\end{gather}

It follows from (\ref{cf18}) that
\begin{gather}\la{cf19}\nu=\frac14,\quad \varphi=\frac{\alpha}{x^2},\quad f_5=\frac{2\alpha}{x^2},\quad f_6=c\end{gather}
where $\alpha$ and $c$ are integration constants.

The next determining equation, i.e., (\ref{e4}), reduces to the following form:
\begin{gather*}(2\nu{F^0}'-\alpha c)\varepsilon^{mab}x^a\nu^b=0,\end{gather*} or
\begin{gather}\frac12{F^0}'=\frac{\alpha c}x.\la{cf20}\end{gather}

On the other hand, equation (\ref{e5}), which  takes the form
\begin{gather*}\nu^axf_7+x^a\nu^cx^cf_7'+(\nu^ax^2-x^a\nu^cx^c){F^0}'=0,
\end{gather*}is solved by the following functions:
\begin{gather}\la{cf21}F^0=f_7=\frac\lambda{x}\end{gather} where $\lambda$ is an integration constant.

Relations  (\ref{cf20}) and (\ref{cf21}) are compatible if either $\alpha=0$ or $\lambda=c=0$. The first possibility corresponds to a trivial vector field $\bf F$ in equation (\ref{Redu}). If $\lambda=c=0$ and $\alpha\neq0$ then formulae (\ref{cf19}) together with (\ref{cf17}), (\ref{cf7}), (\ref{F}) and (\ref{qu}) gives the Hamiltonian and integral of motion presented in equation (\ref{H4}).
\subsubsection{Solutions for functions (\ref{cf5})-(\ref{cf7}) with nontrivial $\Phi^{mab}$}

Functions (\ref{cf5})-(\ref{cf7}) solve the determining equations
(\ref{e1}) and (\ref{e11}). The next step is to substitute
(\ref{cf5})-(\ref{cf7}) and (\ref{F}) into  (\ref{e2})-(\ref{e6}).
Equating in (\ref{e2}) coefficients for linearly independent terms:
\begin{gather*}\la{Y}Y_1^{mab}=(\varepsilon^{mbk}\lambda^a+
\varepsilon^{mak}\lambda^b)x^k,\quad
Y_2^{mab}=2\varepsilon^{mck}\lambda^cx^k\delta^{ab},\quad
Y_3^{mab}=2\varepsilon^{mck}\lambda^cx^kx^ax^b,\\
Y_4^{mab}=(\varepsilon^{mbc}x^a+\varepsilon^{mac}x^b)\lambda^c,\quad
Y_5^{mab}=x^m(x^a\varepsilon^{bcd}+x^b\varepsilon^{acd})x^c\lambda^d\end{gather*}
and using the identities
\begin{gather}\la{cf8}\begin{split}&(x^a\varepsilon^{mbc}+
x^b\varepsilon^{mac})\lambda^c=Y_1^{mab}+Y_4^{mab}-Y_2^{mab},\\&(x^a\varepsilon^{b
mk}+x^b\varepsilon^{a
mk})x^k\lambda^cx^c=Y_3^{mab}+Y_5^{mab}-x^2(Y_1^{mab}+Y_2^{mab}-Y_4^{mab})\end{split}\end{gather}
we obtain the following system of equations for functions $\varphi,
f_1, f_3$ and $f_3$:
\begin{gather}\begin{split}&\nu_4x\varphi+f_2'-f_1'=0,\\
&f_2'-\nu_3x\varphi=0,\\
&\nu_4x^3\varphi+xf_1+x^2f_2'+f_3'=0,\\
&(\nu_3x^2-\nu_1)x\varphi-
x^2f_2'-f_3'=0,\\
&\nu_2x\varphi-x(f_2x)'+f_3'=0.\la{cf9}\end{split}
\end{gather}

The  general solutions for system (\ref{cf9}) look as follows:
\begin{gather}\la{cf10}\begin{split}&\varphi=
\frac\alpha{(x^2+\omega)^\frac32},\quad f_1=-\frac{\alpha(\nu_3+\nu_4)}{(x^2+\omega)^\frac12},\\&
f_2=-\frac{\alpha\nu_3}{(x^2+\omega)^\frac12},\quad f_3=-\frac{\alpha\nu_1}{(x^2+\omega)^\frac12},\quad
\omega=\frac{\nu_1}{\nu_3+\nu_4}\end{split}\end{gather}if $\nu_3(\nu_3+\nu_4)\neq0,\quad \nu_2=(\nu_4+2\nu_3)\omega$, and
\begin{gather}\la{cf11}\varphi=\alpha, \quad f_1=-\alpha\nu_1,\quad f_2=\alpha(\nu_1+\nu_2),\quad f_3=\frac12\alpha\nu_1x^2\end{gather}
if $\nu_3=\nu_4=0$. Here $\alpha$ is an integration constant.

Consider solutions (\ref{cf10}). The next determining equation, i.e., (\ref{e3}), is compatible with (\ref{cf10}) with a nontrivial $\alpha$ only for the case $\omega=0$ and $\nu_1=\nu_2=\nu_3=0$. In accordance with (\ref{cf10}) the corresponding Hamiltonian is reduced to the form given by equation (\ref{H1}). Solving (\ref{e3}) and the remaining equations (\ref{e4}) and (\ref{e5}) we obtain the following second order integral of motion:
\begin{gather}\la{cf13}{\bf Q}={\bf J}\left(\mbox{\boldmath$\sigma$}\cdot{\bf L}+1 +\alpha\mbox{\boldmath
$\sigma$}\cdot{\bf n}\right)\end{gather}
This result is trivial since (\ref{cf13}) is a product of the  known first order integrals of motion for Hamiltonian (\ref{H1}).

Consider the case $\nu_3=\nu_4=0$ and the corresponding solutions (\ref{cf11}).
Substituting  functions (\ref{cf5}), (\ref{cf4}), (\ref{cf7}) and (\ref{cf11}) into (\ref{e3}), and equating coefficients for linearly independent terms, we obtain the following system:
\begin{gather}\la{cf14}\begin{split}&2\nu_0\alpha x-
(2\nu_2+\nu_1)x^3+\nu_2{F^0}'+xf_5=0,\\&f_6'-\nu_0\alpha
x+2\nu_1{F^0}'-\nu_1\alpha^2x^3=0,\\&\nu_2{F^0}'-\nu_0\alpha+xf_5=0,\\&
2(\nu_1+\nu_2)\alpha^2x+f_5'=0.\end{split}\end{gather} This system
has nontrivial solutions iff $\lambda_0=0,\ \nu_1=-2\nu_2$, then
\begin{gather}\la{cf15}f_5=-\nu_2(\alpha^2x^2+2c_1),\quad F^0=\frac{\alpha^2}4x^4+c_1x^2,\quad f_6=\nu_2\left(\frac{g^2}2 x^2+8c_1x^2\right)+c_2\end{gather}
where $c_1$ and $c_2$ are integration constants.

Relations (\ref{cf15}) are compatible with  equation (\ref{e4}) only for $\nu_2\alpha=0$. Thus a nontrivial second order symmetry (which corresponds to $\nu_2\neq0$) is possible for the trivial field $\bf F$.
We show that, except (\ref{cf13}),  there are now second order integrals of motion with even functions (\ref{v1}) provided $\Phi^{mab}$ are nontrivial.
\subsubsection{Solutions for functions (\ref{n1})-(\ref{n3})}
Functions (\ref{n1})-(\ref{n3}) solve the subsystem of the determining equations given by relations (\ref{e1}) and (\ref{e11}). Substituting these functions into the next equation (\ref{e2}) and equating the coefficients for linearly independent terms $x^ax^bx^m\lambda^cx^c,\ x^m\delta^{ab}x^c\lambda^c,\
(\delta^{ma}x^b+\delta^{mb}x^a)x^c\lambda^c, \
\lambda^mx^ax^b,\ x^m(\lambda^ax^b+\lambda^bx^a),\ \lambda^m\delta^{ab}$
and $\delta^{ma}\lambda^b+\delta^{mb}\lambda^a$ we obtain the following system:
\begin{gather}\la{n5}\begin{split}&f_1=\nu_2\varphi,
\quad f_1'=0,\\&f_2'+2f_1x=0,\quad
f_3'+\left(f_1+(\nu_2-\nu_1)\varphi\right)x=0\\&f_4+\nu_2x^2\varphi=0,\quad
f_4-\nu_2x\varphi=0,\\&f_2+f_3+(\nu_1-\nu_2)x^2\varphi=0.
\end{split}\end{gather} Its general solutions are
\begin{gather}\la{bu}f_1=f_4=\nu_2=0,\quad f_2=\mu, \quad f_3=-
\frac{\nu_1\alpha}{x}-\mu,\quad
\varphi=\frac{\alpha}{x^3}\end{gather} where $\alpha$ and $\mu$ are
integration constants.

The next determining equation (\ref{e3}) takes the following form:
\begin{gather*}x^c\lambda^c\varepsilon^{abk}x^k\left(\frac{{F^0}'}{x}+
2\frac{\alpha\mu}{x^3}\right)+\varepsilon^{abk}\lambda^kf_5+
x^a\varepsilon^{bkc}\lambda^kx^c\frac{f_5'}{x}=0.\end{gather*} Its
solutions are $f_5=0,\ F^0=-\frac{2\alpha\mu}x+\text{Const}$.
However, these solutions and solutions (\ref{bu}) are compatible
with equation (\ref{e6}) only for $\alpha=0$, i.e., for the case of the
trivial Pauli interaction.

Thus we prove that vector constants of motion exist only in two
cases, namely,  when the Hamiltonian is given by equation (\ref{H1})
or (\ref{H4}). In the first case integrals of motion are given by
equation (\ref{cf13}) and so are products of the first order
symmetry operators. In the second case the integrals of motion are
components of the Laplace-Runge-Lenz vector with spin, introduced in
\cite{N2}.
\subsection{Tensor integrals of motion}
In accordance with the analysis presented in section A1, the tensors
of the zero and first order in $\frac\p{\p x_a}$ do not exist.
Moreover, to search for second order tensor integrals of motion we
can {\it a priori} restrict ourselves to the case when the
coefficients for second order terms are given either by equation
(\ref{v3}) or by equation (\ref{v4}). Consider both versions
consequently.
\subsubsection{Integrals of motion with odd $\Phi^{0ab}$ and even
$\Phi^{mab}$} The corresponding coefficient functions for the second
order terms are given by equations (\ref{v3}), (\ref{Kt4}) and
(\ref{Ktt4}), where the second order constant tensors should be
proportional:
\begin{gather}\la{Ph1}\Phi^{0ab}=
\nu_1(\lambda^{ac}\varepsilon^{bck}+
\lambda^{bc}\varepsilon^{ack})x^k,\\\begin{split}&\Phi^{mab}=
\nu_2(\varepsilon^{m ac}\lambda^{cb}+\varepsilon^{m
bc}\lambda^{ca})\\&+\nu_3\left((\delta^{m a}\varepsilon^{bkc}+
\delta^{m
b}\varepsilon^{akc})\lambda^{kd}x^cx^d-x^m(\lambda^{ak}\varepsilon^{bkc}
+\lambda^{bk}\varepsilon^{akc})x^c\right).\end{split}\la{Ph2}\end{gather}

The related tensors $\Lambda^{ma}$ and vectors $\Omega^m$ should be
even functions of $\bf x$ linear in tensor parameter $\lambda^{ab}$.
All such functions can be represented in the following form:
\begin{gather}\la{Lamo}\Lambda^{ma}=\lambda^{ma}f_1+
x^m\lambda^{ac}x^cf_2+
x^a\lambda^{mc}x^cf_3+\delta^{ma}\lambda^{cd}x^cx^df_4,\\\la{lamo}\Omega^m=
\varepsilon^{mkc}\lambda^{cb}x^kx^bf_5\end{gather} where $f_1..f_5$
are arbitrary functions of $x$.

The coefficient $\Omega^0$ should be a scalar odd function linear in
$\lambda^{ab}$; such functions do not exist.

Substituting (\ref{Ph2}) (\ref{Lamo}) and (\ref{F}) into (\ref{e2})
and equating coefficients for the linearly independent terms
$x^m\delta^{ab}\lambda^{cd}x^cx^d$,
$\lambda^{ma}x^b+\lambda^{mb}x^a$,
$\delta^{ma}\lambda^{bc}x^c+\delta^{mb}\lambda^{ac}x^c$,
$x^m\lambda^{ab}$, $\delta^{ma}x^b+\delta^{mb}x^a$,
$\delta^{ab}\lambda^{mc}x^c,$ $x^ax^b\lambda^{mc}x_c$ and
$x^m(x^b\lambda^{ac}+x^a\lambda^{bc})$, we obtain the following
conditions:
\begin{gather}\begin{split}&\nu_3=0,\quad f_2=0,\quad f_3=0,\quad
f_4=\mu,\\&f_1'+\nu_2x\varphi=0,\quad
2f_4-\nu_2\varphi=0\end{split}\la{ee21}\end{gather} where $\mu$ is
an integration constant.

 Let $\nu_2\neq0$ then it follows from
(\ref{ee21}) that
\begin{gather}\la{coco}F^\mu=\frac{2\mu x^m}{\nu_2},\quad \Lambda^{m
b}=\mu(\delta^{mb}\lambda^{cd}x^cx^d-\lambda^{mb}x^2). \end{gather}

If $\nu_2=0$ then $f_4=0,\ f_1=\tilde\mu$, and so
\begin{gather}\Phi^{mab}=0,\quad \Lambda^{m
b}=\tilde\mu\lambda^{mb},\quad F^m=x^m\varphi\la{cococo}\end{gather}
where $\tilde\mu$ is a constant and $\varphi$ is an arbitrary
function.

Substituting (\ref{Ph1}),  (\ref{coco}) and (\ref{Ph2}) with
$\nu_3=0$ into the next determining equation (\ref{e3}) and equating
to zero the coefficient for the term
$\varepsilon^{mak}x^k\lambda^{cd}x^cx^d$ we obtain the condition
$\mu=0$. In accordance with (\ref{coco}) it corresponds to the
trivial external vector field and so to the trivial Pauli term in
Hamiltonian (\ref{Redu}).

Consider now the version represented by equations (\ref{cococo}).
Substituting (\ref{Ph1}) and (\ref{cococo}) into (\ref{e3}), using
the identities
\begin{gather}\begin{split}&\varepsilon^{ack}\lambda^{cm}x^k=
\varepsilon^{mck}\lambda^{ca}x^k+\varepsilon^{mac}\lambda^{ck}x^k,\\&
\varepsilon^{mak}x^k\lambda^{cd}x^cx^d=x^m\varepsilon^{akc}x^k\lambda^{cb}x^b
-x^a\varepsilon^{mkc}x^k\lambda^{cb}x^b
-x^2\varepsilon^{mac}\lambda^{ck}x^k\end{split}\la{iden}\end{gather}
and equating the coefficients for linearly independent terms
presented in the r.h.s. of equation (\ref{iden}) we obtain the
following conditions:
\begin{gather}\varphi'=0,\quad f_5'=0,\quad 2\nu_1\varphi-f_5=0,\quad
\nu_1\varphi-2\tilde\mu\varphi+f_5=0.\la{kukur}
\end{gather}

It follows from (\ref{kukur}) that
\begin{gather}\la{kukuri}
\varphi=\nu,\quad f_5=2\nu\nu_1,\quad
\tilde\mu=\frac32\nu_1\end{gather} where $\nu$ is a constant.
However, equations (\ref{lamo}), (\ref{cococo}) and (\ref{kukuri})
are incompatible with (\ref{e4}) if $\varphi=\nu\neq0$, thus there
are no tensor integrals of motion of the considered type for
Hamiltonian (\ref{Redu}) with a non-trivial matrix term
$\mbox{\boldmath $\sigma$}\cdot {\bf F}$.

\subsubsection{Integrals of motion with even $\Phi^{0ab}$ and odd
$\Phi^{mab}$} The related coefficient functions for the second order
terms are given by equations (\ref{v4}), (\ref{Kt3}) and
(\ref{Ktt5}). Moreover, without loss of generality it is possible to
set in (\ref{Kt3}) $\lambda_2^{ab}=0$ since the terms including this
coefficient can be related to the obvious symmetry operators
$\lambda_2^{ab}J_aJ_b$ where $J_a$ and $J_b$ are components of the
total orbital momentum. Thus we have:
\begin{gather}\la{t1}\Phi^{0 a b}=\nu_1\lambda^{ ab}
\\\la{t2}\Phi^{m a b}=\nu_2(
\lambda^{m a}x^b+\lambda^{m b}x^a-2\delta^{ab}\lambda^{m
c}x_c)+\nu_3(2x^m\lambda^{ab}-
(\delta^{ma}\lambda^{bc}+\delta^{mb}\lambda^{ac})x_c).\end{gather}

The corresponding tensors $\Lambda^{ma}$ and vectors $\Omega^m$
should be odd functions of $\bf x$ linear in tensor parameter
$\lambda^{ab}$. The generic form of such functions is given by the
following equations:
\begin{gather}\la{t3}\Lambda^{mb}=\varepsilon^{mcb}\lambda^{ck}x^kf_1+
\varepsilon^{bck}\lambda^{mk}f_2+\varepsilon^{mbc}x^c\lambda^{dk}x^dx^kf_3+
\varepsilon^{mkc}\lambda^{kd}x^bx^cx^df_4,\\\la{t4}
\Omega^a=\lambda^{ac}x^cf_5+x^a\lambda^{kd}x^kx^df_6\end{gather}where
$f_1..f_6$ are functions of $x$.

In addition, there is a possible odd non-derivative term $\Omega^0$
in (\ref{qu}):
\begin{gather}\la{t5}\Omega^0=\lambda^{kd}x^kx^df_7.\end{gather}

Substituting (\ref{t2}) and (\ref{t3}) into (\ref{e2}) we obtain the
following equation:
\begin{gather}\begin{split}&X_3^{mab}\frac{f_2'}x+X_8^{mab}\frac{f_1}{x^2}+
X_2^{mab}\frac{f_1'}x+2X_7^{mab}f_3+X_9^{mab}\frac{f_3'}x+X_6^{mab}f_4\\&-
(X_1^{mab}+X_2^{mab})f_4+X_{10}^{mab}\frac{f_4'}x-(\nu_2X_7^{mab}
-\nu_3X_3^{mab}+\nu_3X_5^{mab})\varphi=0\end{split}\la{t6}\end{gather}
where
\begin{gather}\begin{split}&X_1^{mab}=(x^a\varepsilon^{mcb}+x^b\varepsilon^{mca})
\lambda^{ck}x^k,\quad
X_2^{mab}=(x^a\lambda^{bc}+x^b\lambda^{ac})\varepsilon^{mck}x^k,\\&
X_3^{mab}=\lambda^{mk}(x^a\varepsilon^{bck}+x^b\varepsilon^{ack})x^c,\quad
X_4^{mab}=x^m(\lambda^{ak}\varepsilon^{bck}+
\lambda^{bk}\varepsilon^{ack})x^c,\\&X_5^{mab}=(\delta^{ma}\varepsilon^{bdk}
+\delta^{mb}\varepsilon^{adk})x^d\lambda^{kc}x^c,\quad
X_6^{mab}=2\delta^{ab}\varepsilon^{mck})x^c\lambda^{kd}x^d,\\&
X_7^{mab}=(\varepsilon^{mbc}\lambda^{ak}+\varepsilon^{mac}\lambda^{bk})x^cx^k,
\quad
X_8^{mab}=(\varepsilon^{mcb}\lambda^{ac}+\varepsilon^{mca}\lambda^{bc})x^2,\\
&X_9^{mab}=(x^a\varepsilon^{mbc}+x^b\varepsilon^{mac})\lambda^{kd}x^cx^kx^d,\quad
X_{10}^{mab}=2x^ax^b\varepsilon{mck}x^c\lambda^{kd}x^d.\end{split}\la{t7}\end{gather}

Tensors (\ref{t7}) satisfy the following conditions:
\begin{gather}\begin{split}&X_{3}^{mab}=X_{1}^{mab}-X_{3}^{mab},\\&X_{4}^{mab}=
X_{7}^{mab}+X_{8}^{mab}-X_{2}^{mab},\\& X_{5}^{mab}=X_{1}^{mab}+
X_{6}^{mab}+X_{7}^{mab}\end{split}\la{t8}\end{gather} while all
terms in the r.h.s. of (\ref{t8}) are linearly independent. Thus
equation (\ref{t6}) generates the following system of equations:
\begin{gather}\la{t9}\begin{split}&f_1=0,\quad
f_3'=f_4'=0,\\&2f_3=(\nu_2+\nu_3)\varphi,\quad
f_4=\nu_3\varphi,\\&f_2=f_4x,\quad f_2'+f_4x+\nu_3\varphi
x=0.\end{split}\end{gather}Solutions of this system with a
non-trivial $\varphi$ are:
\begin{gather}f_1=f_4=\nu_3=0,\quad f_2=\mu,\quad
f_3=\frac{\nu_2\lambda}2,\quad \varphi=\lambda\la{t10}\end{gather}
where $\mu$ and $\lambda$ are arbitrary constants.

In accordance with (\ref{t10}) coefficient functions (\ref{t3}) and
vector field $F^a$ are reduced to the following form:
\begin{gather}\la{t11}\Lambda^{\nu
a}=\mu\varepsilon^{ack}x^c\lambda^{\nu
k}+\frac12\nu_2\lambda\varepsilon^{\nu ac}\lambda^{d
k}x^cx^dx^k,\quad F^a=\lambda x^a.\end{gather}

Substituting (\ref{t1}), (\ref{t2}) with $\nu_3=0$, (\ref{t4}) and
(\ref{t11}) into (\ref{e3}) and equating the coefficients for
linearly independent terms $\lambda^{am},
\delta^{am}\lambda^{cd}x^cx^d, x^ax^m, x^m\lambda^{ak}x^k$ and
$x^a\lambda^{mk}x^k$ we obtain the following system:
\begin{gather*}2\mu\lambda=0,\\\nu_1\lambda+f_5=0,\\f_6+\nu_2\lambda^2x^2-\frac{\nu_2}x {F^0}'=0,\\
f_6'-\nu_2\lambda^2x=0,\\ \nu_2{F^0}'+f_5'+2f_6x=0.\end{gather*}

This system is compatible only for $\nu_2=0$ or $\lambda=0$. The
first case is compatible with (\ref{e4}) only for $\lambda=0$, thus
all admissible  solutions correspond to a trivial external vector
field $F^a$.

Summarizing, there are no rank 2 tensor integrals of motion for equation
(\ref{Redu}) if ${\bf F} \neq0$.

Integrals of motion which are tensors of rank 3 can be excluded a priory since they have too many independent components whose number exceeds the maximal possible number of constants of motion for a 3d system. The absence of such integrals of motion (and integrals being tensors of rank R$>$3) also can be proven directly using the determining equations (\ref{e1})-(\ref{e6}).

\end{document}